\documentclass[prl,twocolumn,showpacs,preprintnumbers,amsmath,amssymb]{revtex4}
\usepackage{graphicx}% Include figure files with \includegraphics
\usepackage{epsfig}% Include figure files with \psfig
\newcommand{\be}{\begin{eqnarray}}
\newcommand{\ee}{\end{eqnarray}}

\newcommand{\bi}{\begin{itemize}}
\newcommand{\ei}{\end{itemize}}

\begin{document}
%\twocolumn[\hsize\textwidth\columnwidth\hsize
%           \csname @twocolumnfalse\endcsname
\title{Vortex induced deformation of the superconductor crystal lattice}
\author{Pavel Lipavsk\'y$^1$, Klaus Morawetz$^{2,3}$, 
Jan Kol\'a\v cek$^4$ and Ernst Helmut Brandt$^5$}
\affiliation{
$^1$ Faculty of Mathematics and Physics, Charles University, 
Ke Karlovu 3, 12116 Prague 2, Czech Republic}
\affiliation{
$^2$Institute of Physics, Chemnitz University of Technology, 
09107 Chemnitz, Germany}
\affiliation{
$^3$Max-Planck-Institute for the Physics of Complex
Systems, Noethnitzer Str. 38, 01187 Dresden, Germany}
\affiliation{
$^4$Institute of Physics, Academy of Sciences, 
Cukrovarnick\'a 10, 16253 Prague 6, Czech Republic}
\affiliation{
$^5$Max-Planck-Institute for Metals Research,
         D-70506 Stuttgart, Germany}
\begin{abstract}
Deformation of the superconductor crystal lattice caused 
by Abrikosov vortices is formulated as a response of the elastic 
crystal lattice to electrostatic forces. It is shown that the lattice 
compression is linearly proportional to the electrostatic 
potential known as the Bernoulli potential. 
Eventual consequences of the crystal lattice deformation 
on the effective vortex mass are discussed.
\end{abstract}

\pacs{
%574.20.De      %Phenomenological theories (two-fluid, Ginzburg-Landau, etc.)
74.25.Fy, %      Transport properties (electric and thermal conductivity, thermoelectric effects, etc.)
74.25.Ld, %     Mechanical and acoustical properties, elasticity, and ultrasonic attenuation
74.25.Qt, %     Vortex lattices, flux pinning, flux creep
,74.81.-g, %     Inhomogeneous superconductors and superconducting systems
%03.65.Nk %      Scattering theory
%,21.45.+v %     Few-body systems
%,72.10.Fk %     Scattering by point defects, dislocations, surfaces, and other imperfections (including Kondo effect)
%,03.65.Ge %     Solutions of wave equations: bound states
%,34.80.Pa %     Coherence and correlation in electron scattering
%,34.10.+x %     General theories and models of atomic and molecular collisions and interactions (including statistical theories, transition state, stochastic and trajectory models, etc.)
%,68.65.Hb %Quantum dots
%,73.22.-f %     Electronic structure of nanoscale materials: clusters, nanoparticles, nanotubes, and nanocrystals
%,79.20.Rf %    Atomic, molecular, and ion beam impact and interactions with surfaces
%, 61.14.Dc %    Theories of diffraction and scattering
%,61.46.+w %     Nanoscale materials: clusters, nanoparticles, nanotubes, and nanocrystals
%71.45.Gm, %      Exchange, correlation, dielectric and magnetic response functions, plasmons
78.20.-e, %      Optical properties of bulk materials and thin films 
%78.47.+p, %      Time-resolved optical spectroscopies and other ultrafast optical measurements in condensed matter 
%42.65.Re, %     Ultrafast processes; optical pulse generation and pulse compression
%82.53.Mj %     Femtosecond probing of semiconductor nanostructures 
}
\maketitle
%    \vskip2pc]

During the transition from a normal to a superconducting state, 
metals reduce their specific volumes \cite{S52,H68}. In a mixed state,
would it be the Abrikosov vortex lattice or a structure of lamellas, 
the superconductivity is locally suppressed and the specific
volume is inhomogeneous. The mixed state is thus accompanied by
strains and stresses, which enter the balance of total energy.

In general, the energy of strains is much smaller than the 
energy of the superconducting condensation and the energy of
the magnetic field. Its contribution becomes appreciable only
under special conditions. For example, experiments on single 
crystals of Pb-alloys \cite{O71,SUS71} and Nb-alloys 
\cite{SUS71,E71,WSL73} revealed that an orientation of the 
vortex lattice is influenced by its angles to main crystal 
axes. Since the gap of alloyed samples is quite isotropic,
purely electronic models have failed and this effect has been 
explained with the help of strains induced by 
vortices \cite{UZD73}.
 
In the 1990th a different structural effect has been observed
on NbSe$_2$. If the magnetic field is tilted from the $c$-axes, 
the Ginzburg-Landau (GL) theory predicts a state in which rows 
of vortices are aligned with the direction of tilting \cite{CDK88},
while experiments \cite{HMW92,HMW94,BCGWB93,GHKBOBBMN94} show
them aligned in the perpendicular direction. Again, the 
interaction of vortices with the crystal strain explains the 
observed alignment \cite{KBMD95}.

Finally, we would like to mention phenomena which 
are predicted but not yet fully experimentally confirmed. 
Perhaps, one of the most interesting predictions is a sizable 
contribution of the lattice deformation to the mass of vortex 
\cite{S91,DS92,C94,CK03}. Besides, there is a number of 
phenomena due to strains at surfaces which are discussed 
in Ref.~\cite{SF89}. It is also argued that the strain 
can mediate an attractive long-range interaction between 
vortices \cite{CLM03}.

As far as we know, all theoretical studies of deformable
superconductors use a phenomenological model, which assumes
that the superconducting condensate interacts with the 
lattice density directly via a strain dependence of 
material parameters. This model dates back to the 1960th 
\cite{KB67,L68}, when it was used to describe the vortex 
pinning. The strength of the interaction is deduced from 
changes of the specific volume in the phase transition, 
see also Ref.~\cite{S91}. 

In this paper we assume that the condensate interacts with 
the crystal lattice via electrostatic forces created by the 
so called Bernoulli potential. We show that this 
mechanism results in the interaction based on the specific 
volume. In addition to known theories we obtain gradient corrections 
and demonstrate that they are important for the vortex motion of the 
Abrikosov vortex lattice in niobium. 

In an isotropic continuum, the displacement field $\bf u$ obeys
the equation \cite{LL75}
\begin{equation}    
\left(K+{4\over 3}\mu\right)\nabla(\nabla.{\bf u})-
\mu\nabla\times\nabla\times{\bf u}={\bf F},
\label{elast}
\end{equation}
where $K$ and $\mu$ are the bulk and shear modulus, and
$\bf F$ is the volume density of force acting on the lattice.
Some authors prefer to express coefficients on the left
hand side in terms of the Poisson ratio
$\sigma=(3K-2\mu)/(6K+2\mu)$ and the Yang modulus
$E=3K(1-2\sigma)$. In the basic approximation $K$ and $\mu$
are constants. Their small change in the superconducting 
state was assumed in \cite{K69}.

The inhomogeneous superconductivity results in the force
$\bf F$. The present theory differs from previous approaches 
in the approximation adopted for $\bf F$. Let us sketch 
the approach based on the specific volume first. More
details the reader can find in Ref.~\cite{DS92}.

In the phase transition from the normal to the superconducting 
state, a system shrinks by a volume difference $\delta V=
V_{\rm n}-V_{\rm s}=\alpha_T\, V$. A typical value of 
$\alpha_T$ is about $10^{-7}$. The density of the atomic 
lattice correspondingly increases, $\delta n_{\rm lat}=
n_{\rm lat}^{\rm s}-n_{\rm lat}^{\rm n}=\alpha_T\, n$.

The strain coefficient $\alpha_T$ depends on the temperature
via a fraction $\omega$ of electrons, which become 
superconducting, $\alpha_T=\alpha\omega$. Here $\alpha$ is
the strain coefficient at zero temperature. In the spirit 
of the GL theory, we express the superconducting fraction 
in terms of the GL function $\omega=2|\psi|^2/n$, i.e., 
$\delta n_{\rm lat}=2\alpha\,|\psi|^2$. 

\begin{figure}%[h]   % 1
\centerline{\parbox[c]{8cm}{
\psfig{figure=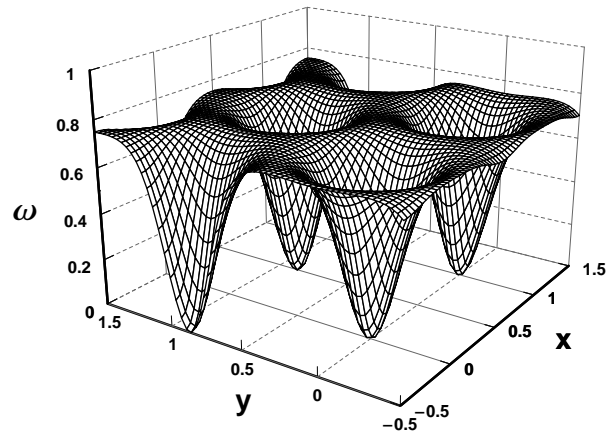,width=8cm}}}
\vskip -5mm
\caption{The superconducting fraction $\omega=2|\psi|^2/n$ 
in the triangular 
Abrikosov vortex lattice. We assume niobium with the
GL parameter increased by non-magnetic impurities to 
$\kappa=1.5$, the temperature $T=0.7\,T_{\rm c}$ and the 
mean magnetic field $\bar B=0.24\,B_{\rm c2}$. In 
centers of vortices the superconducting fraction 
$\omega$ goes to zero, while between vortices it 
approaches its non-magnetic value $1-T^4/T_{\rm c}^4=0.76$.}
\label{fig1}
\end{figure}

In the vortex core or in the region of surface currents, 
the GL wave function changes in space, see Fig.~\ref{fig1}. 
The lattice then tends to be inhomogeneous, which causes 
internal stresses. These stresses lead to a density of force
\begin{equation}    
{\bf F}_{\rm Sim}=K\,\alpha\,\nabla{2|\psi|^2\over n}
\label{forces}
\end{equation}
proposed by Duan and {\v S}im{\'a}nek \cite{DS92} in 
their study of the vortex mass. The phenomenological 
formula (\ref{forces}) furnishes us with a density of 
forces with no regards to how the superconducting
electrons are coupled to the lattice.

Let us try to express 
such force on the lattice in a semi-microscopic way. Diamagnetic 
currents, either being on the surface or circulating around 
the vortex core, always cause inertial and Lorentz forces,
which are balanced by an electrostatic field ${\bf E}=-\nabla
\phi$, see London \cite{L50}. This electric field transfers
the Lorentz force from electrons to the lattice, therefore one 
can expect that it also causes lattice deformations. 
Accordingly, we suppose that the electrostatic field force 
\begin{equation}    
{\bf F}=en\,\nabla\phi,
\label{forcephi}
\end{equation}
is playing the role of the force ${\bf F}$ in equation (\ref{elast}).
For simplicity of notation we assume singly ionized atoms, 
$e$ is the charge of an electron, so that the ionic charge
density is $-en$. 

\begin{figure} [b]   % 1
\centerline{\parbox[c]{8cm}{
\psfig{figure=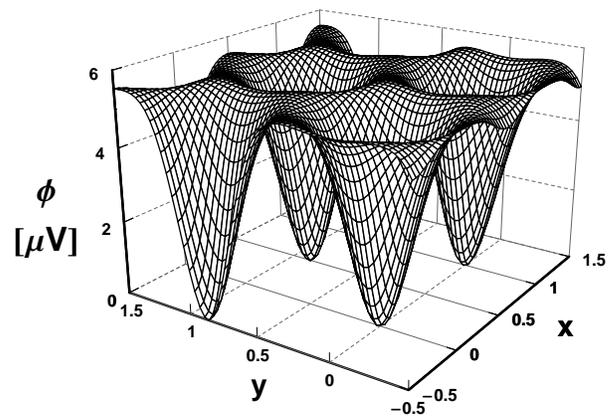,width=8cm}}}
\vskip -5mm
\caption{The electrostatic potential for the parameters of 
Fig.~\protect\ref{fig1}. Its shape reminds the 
superconducting fraction shown in Fig.~\protect\ref{fig1},
which indicates that corrections beyond the 
approximation of Khomskii and Kusmartsev are small.
}
\label{fig2}
\end{figure}

\begin{figure}%[h]   % 1
\centerline{\parbox[c]{8cm}{
\psfig{figure=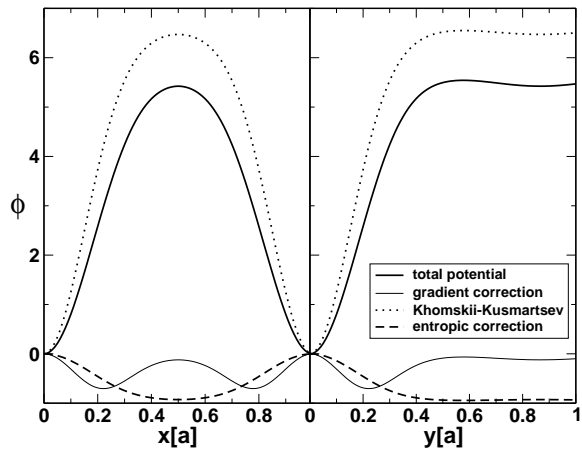,width=8cm}}}
\caption{Cuts through the electrostatic potential from 
Fig.~\protect\ref{fig2}. The thick full line represents 
the total potential (\ref{phi}), the dot line is 
the Khomskii-Kusmartsev approximation. The thin full line
is the gradient correction given by the `kinetic energy'
term of (\ref{phi}), and the dashed line is the 
entropic correction given by the last term of (\ref{phi}). 
}
\label{fig3}
\end{figure}

The electrostatic potential $\phi$ is known as the 
Bernoulli potential. It has been derived in a number 
of approximations \cite{L50,VS64,KK92,LKMB02}. Here we 
will use the formula of Ref.~\cite{LKMB02}
\begin{eqnarray} % 59
e\phi&=&-{1\over 2m^*n}
\bar\psi\left(-i\hbar\nabla-e^*{\bf A}\right)^2\psi
\nonumber\\
&+&{\partial\varepsilon_{\rm con}\over\partial n}
{2|\psi|^2\over n}+{T^2\over 2}{\partial\gamma\over\partial n}
\left(\sqrt{1-{2|\psi|^2\over n}}-1\right).
\label{phi}
\end{eqnarray}
The space profile of the potential $\phi$ is shown in 
Fig.~\ref{fig2} and the individual terms to the potential are 
compared in Fig.~\ref{fig3}.
The first term in (\ref{phi}) is the quantum kinetic energy and represents 
the gradient corrections. In the
London limit it reaches the form of the classical 
Bernoulli law, $\bar\psi\left(-i\hbar\nabla-e^*
{\bf A}\right)^2\psi/2m^*n\to e^{*2}A^2\omega/4m^*=
\omega mv^2/2$, which gave the name to the entire potential.
The superconducting fraction $\omega$ multiplying the kinetic 
energy accounts for the fact that the Lorentz and inertial 
forces act exclusively on the super-electrons, while the 
balancing electrostatic force acts on all electrons \cite{VS64}.
This force being proportional to the square of the magnetic field 
has been used to calculate the shape distortion by 
flux-pinning-induced magnetostriction \cite{JLB98}. 

The second term of the Bernoulli potential (\ref{phi})
is the dominant one and we will focus our discussion on
it. This second term is identical to the potential derived 
by Khomskii and Kusmartsev \cite{KK92} from the effect of 
the BCS gap on the local density of electronic states.
The third term of (\ref{phi}) we call the entropic 
correction.

It is worth noting that all the three equations (\ref{phi}), 
(\ref{forcephi}) and (\ref{elast}) are resulting from Gibbs 
variational principle if the free energy used in \cite{LKMB02} 
is extended by the  ionic lattice deformation energy. 

Let us link the deformation caused by the electrostatic
field with the standard theory of magnetostriction.
At zero temperature, the Gibbs energy of normal 
and superconducting states differ by the condensation
energy \cite{H68}, $G_{\rm s}=G_{\rm n}-V
\varepsilon_{\rm con}$, where $\varepsilon_{\rm con}=
\gamma T_{\rm c}^2/4=B_{\rm c}^2/2\mu_0$. Since the 
pressure derivative 
of the Gibbs energy determines the sample volume, 
$V_{\rm s,n}=\partial G_{\rm s,n}/\partial p$, one finds 
$V_{\rm s}=V_{\rm n}-V\,\partial\varepsilon_{\rm con}/
\partial p$, i.e., $\alpha=\partial\varepsilon_{\rm con}/
\partial p$. This relation allows us to show that the 
force ${\bf F}_{\rm Sim}$ from (\ref{forces}) equals 
the electrostatic force caused by the second term of the 
Bernoulli potential (\ref{phi}). 

To proceed, we will use the fact that the pressure modifies the 
condensation energy indirectly by an increase of the 
electron density
\begin{equation}    
\alpha={\partial\varepsilon_{\rm con}\over\partial p}=
{\partial\varepsilon_{\rm con}\over\partial n}
{\partial n\over\partial p}=
{\partial\varepsilon_{\rm con}\over\partial n}
{n\over K}.
\label{alpha}
\end{equation}
In the rearrangement we have employed the definition
of the bulk modulus $K=-V/(\partial V/\partial p)=n/
(\partial n/\partial p)$. Substituting (\ref{alpha}) 
into the force (\ref{forces}) one finds
\begin{equation}    
{\bf F}_{\rm Sim}=n\,
{\partial\varepsilon_{\rm con}\over\partial n}~
\nabla{2|\psi|^2\over n}.
\label{forcesub}
\end{equation}

Comparing (\ref{forcesub}) with (\ref{forcephi}) one can
see that the force introduced by {\v S}im{\'a}nek equals
to the electrostatic force due to the second term of the
Bernoulli potential (\ref{phi}). In this sense, the 
approximation used within the theory of deformable 
superconductors is equivalent to the approximation of
the electrostatic potential derived by Khomskii and 
Kusmartsev. The gradient and entropic corrections of the 
Bernoulli potential (\ref{phi}) provide us with corresponding 
corrections to the force of {\v S}im{\'a}nek.

More interesting is the gradient correction. From the
formula (\ref{phi}) follows that far from the vortex 
core, where $\bar\psi\psi\to (1-T^4/T_{\rm c}^4)\,n/2$, 
the gradient correction is proportional to the square 
of the local current. For an isolated vortex, the 
gradient correction thus decays on the scale of the 
London penetration depth. Consequently one can expect, that it
plays an important role in high $\kappa$ materials. 

The effect of the gradient correction is traceable also
for a conventional material assumed here. To be specific,
our sample is a niobium rod parallel to the magnetic 
field. The GL coherence length of niobium is reduced 
by non-magnetic impurities so that the GL parameter is 
increased to $\kappa=1.5$, while other material 
parameters remain close to values of the pure niobium. 
All plots are for $T=0.7\,T_{\rm c}$ and $\bar B=0.24
\,B_{\rm c2}$. In this case the magnetic field is not 
split into separated unitary fluxes but it is nearly 
homogeneous with amplitude fluctuations of about 20\% 
around the mean field $\bar B$. The vortex cores are well 
separated, however, as $\omega$ reaches its non-magnetic 
value in the out-of-core region, see Fig.~\ref{fig1}.

Let us analyze the three contributions to the 
electrostatic potential from the point of view
of forces they cause. According to the position of the
inflex point of the Khomskii-Kusmartsev potential seen
in Fig.~\ref{fig3}, one can estimate that the maximum 
of the {\v S}im{\'a}nek 
force is at about $x,y\sim 0.1\,a$. This is quite close 
to the center of the vortex core. The gradient correction 
oscillates fast in space having a magnitude much smaller
than the Khomskii-Kusmartsev potential. Its gradient, 
however, is rather comparable to the gradient of the 
dominant term, which shows that gradient contribution
to the force can appreciably modify the {\v S}im{\'a}nek 
force. In the heart of the vortex core, the gradient 
correction to the force acts against the {\v S}im{\'a}nek 
force, while in the skin of the vortex core it points in 
the same direction. As a result, the maximum of the total 
force is shifted outwards to $x,y\sim 0.2\,a$.

\begin{figure}%[h]   % 1
\centerline{\parbox[c]{8cm}{
\psfig{figure=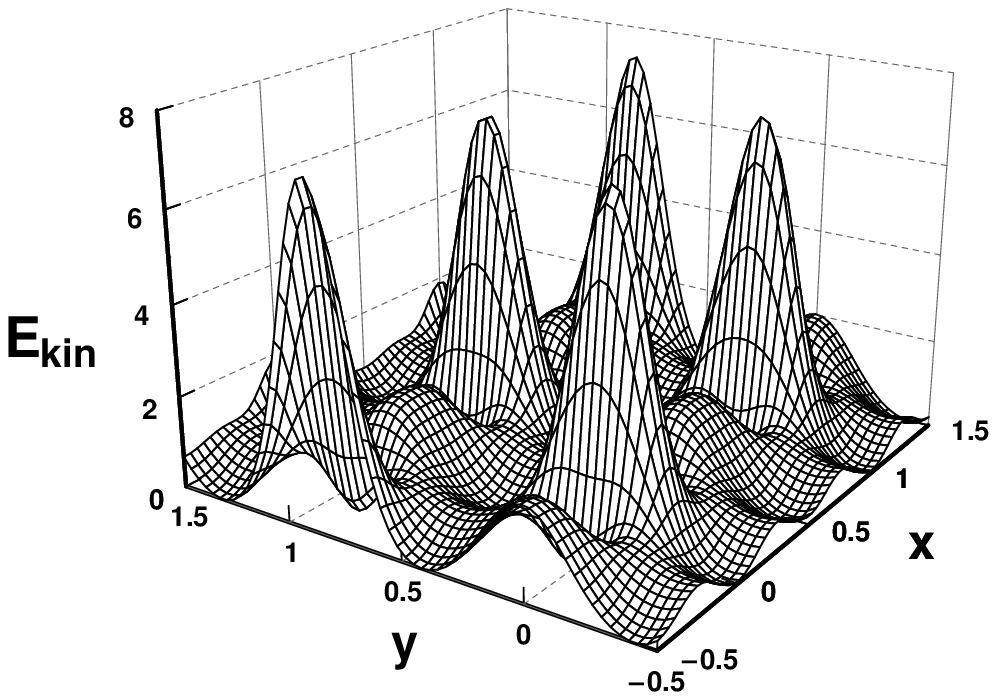,width=8cm}
\vskip -1cm
\psfig{figure=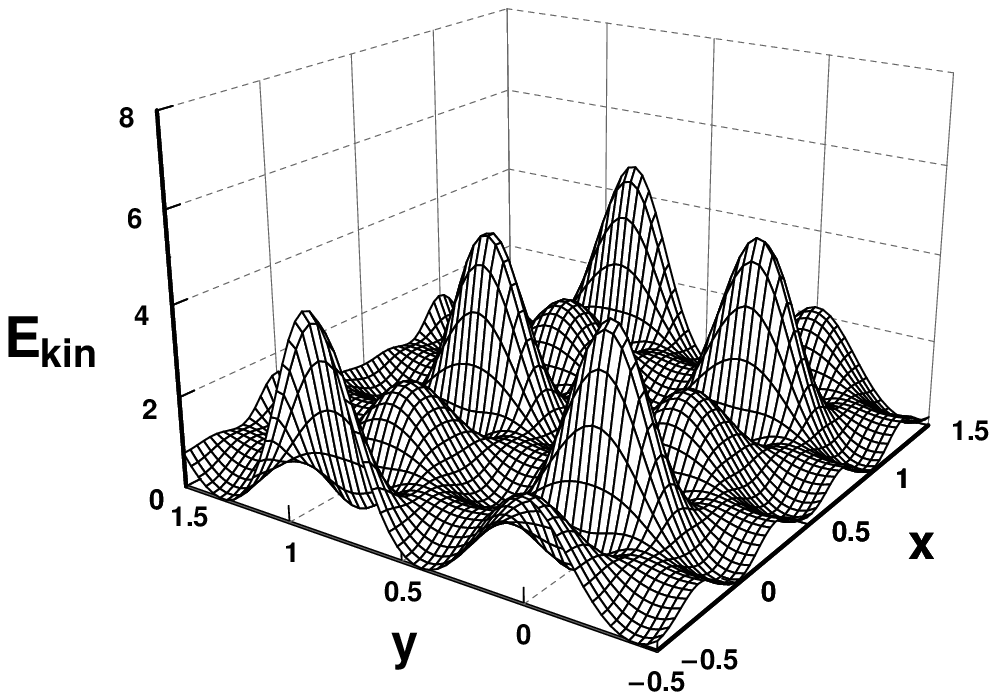,width=8cm}
}}
\caption{The density of kinetic energy of lattice ions 
created by vortices moving in the $x$ 
direction for parameters of Fig.~\protect\ref{fig1}.
The kinetic energy due to (a) {\v S}im{\'a}nek force 
(\ref{forcesub}) , and (b) the electrostatic force 
(\ref{forcephi}). The plotted function is dimensionless
a quantity identical to the bracket from the formula 
(\ref{kinendens}).
} 
\label{fig4}
\end{figure}

Naturally, gradient corrections modify the vortex mass.
Figure~\ref{fig4}a shows how individual regions in the 
Abrikosov vortex lattice contribute to the vortex mass 
for the {\v S}im{\'a}nek force, in figure~\ref{fig4}b 
the gradient and the entropic corrections are included. The 
plotted function is the density of kinetic energy of 
lattice ions driven by vortices moving with velocity 
$V$ in the $x$ direction \cite{S91}
\begin{equation}    
E_{\rm kin}={1\over 2}V^2~n\,M 
\left [\left({\partial u_x\over\partial x}\right)^2+
\left({\partial u_y\over\partial x}\right)^2\right ],
\label{kinendens}
\end{equation}
where $M$ is the mass of a single ion.

In most places of the Abrikosov vortex lattice, the 
ionic kinetic energy is lowered by the correction terms,
see Fig.~\ref{fig4}. From the integral over the 
elementary cell one obtains the vortex mass per unit
length. Keeping both corrections, the ion contribution 
to the vortex mass is reduced by a factor 0.83 as 
compared to {\v S}im{\'a}nek result. If one neglects 
the gradient correction keeping only the entropic 
correction, the reduction factor is 0.70. The gradient 
correction thus leads to a small enhancement of 
the vortex mass.

Comparing relative amplitudes of the kinetic energy 
inside and between the cores, one can see that the 
corrections have increased the share of the out-of-core 
region. Since the entropy term merely reduces 
the amplitude of the {\v S}im{\'a}nek force, this 
redistribution of distortions is exclusively due to
the gradient correction. According to Cano, Levanyuk 
and Minyukov \cite{CLM03} the out-of-core lattice 
deformations are important for the strain-mediated 
interaction of vortices. We suggest that the theory of
this interaction should be reexamined with the gradient
correction included.

In summary, we have expressed the forces deforming a lattice
of a superconductor in terms of the electrostatic force.
This approach allows one to benefit from experiences 
accumulated within the theory of the so called Bernoulli 
potential. One directly obtains the gradient 
corrections we have discussed in this paper. We suggest that the
theory of vortex motion should be revised taking this gradient corrections into account.
The present theory is restricted to homogeneous isotropic 
materials. Its extension to layered materials we plan to discuss 
in short future.

\medskip
This work was supported by research plans
MSM 0021620834 and No. AVOZ10100521, by grants
GA\v{C}R 202/04/0585 and 202/05/0173, and by European 
ESF program AQDJJ.

\end{document}